\documentclass[conference]{IEEEtran}
\ifCLASSINFOpdf
  \usepackage[pdftex]{graphicx}
  \DeclareGraphicsExtensions{.pdf,.jpeg,.png}
\else
\fi

\usepackage[tight,footnotesize]{subfigure}
\usepackage{url}
\usepackage{pifont}
\usepackage{tabulary}

\ifCLASSINFOpdf
\else
\fi
\begin{document}
\title{On Web User Tracking: How Third-Party Http Requests Track Users' Browsing Patterns for Personalised Advertising}

\author{\IEEEauthorblockN{Silvia Puglisi, David Rebollo-Monedero and Jordi Forn\'e}
\IEEEauthorblockA{Department of Telematics Engineering, \\ 
Universitat Polit\`ecnica de Catalunya (UPC)\\
C.\ Jordi Girona 1-3, 08034 Barcelona, Spain\\
silvia.puglisi@upc.edu\\
david.rebollo@entel.upc.edu\\
jforne@entel.upc.edu
}}


%Telephone: (800) 555--1212, Fax: (888) 555--1212}


\maketitle

\begin{abstract}
On today's Web, users trade access to their private data for content and services. Advertising sustains the business model of many websites and applications. Efficient and successful advertising relies on predicting users' actions and tastes to suggest a range of products to buy. It follows that, while surfing the Web users leave traces regarding their identity in the form of activity patterns and unstructured data. We analyse how advertising networks build user footprints and how the suggested advertising reacts to changes in the user behaviour.
\end{abstract}


% \fi
\IEEEpeerreviewmaketitle

\section{Introduction}
Websites use \emph{personalisation services} to profile their visitors, collect their in page reading activities and eventually use this data to provide tailored suggestions. Among the data analysed by websites are also included user preferences and social connections. These can be obtained by tracking users across different applications and sites through cookies or open web sessions. Even if the user does not accept cookies or is not logged into a service account, such as their Google, Twitter or Facebook accounts, the web page and third party services can still try to profile them by using third-party http requests, among other techniques. Within the http request various selectors can be included to communicate user preferences or particular features, in the form of url variables. Features that might be used by advertising networks and malicious trackers include personalised language or fonts settings, browser extensions, in page keywords and so on. These features are then used to identify individual users by restricting the pool of possible candidates among all the visitors in a certain time frame. Unique users can then be distinguished across multiple devices or sessions.

\subsection{Contribution}

We have observed how users are tracked across the Web and how the displayed advertising is tailored even after they have visited a few websites with a certain interest bias. In our study we analyse how the user profile detected by advertising services can be used to estimate the user privacy risk on a certain network. In our study we analyse how advertising networks identify a user and start tracking them. We measure the distance between the observed user profile and the advertising profile, by categorising the set of keywords sent by advertising networks through third-party http requests. We introduce a set of metrics to express this distance between the two profiles. 

It is important to note that we have considered the case for which users are not registering, neither connecting any external account, as it could be the case with services like: Facebook, Google+, Twitter, and so on. In such scenario we have measured how these networks still attempt to track the user by sending user information through http requests to their services.

The main contributions of this paper are the following.
\begin{enumerate}
 \item An analysis of how tracking happens on the web, for users who are not logged in and based on real browsing patterns, taking as sample Google services.
 \item A model of the user online footprint that is able to expose how each website and tracking network categorise their activities.
 \item A measure of how each website affects the advertising returned from advertising services.
 \item A measure of connectivity of malicious trackers across different websites.
\end{enumerate}

\section{Background}

Information regarding locations, browsing habits, communication records, health information, financial information, and general preferences regarding user online and offline activities are shared by different parties online. This level of access is often directly granted from the user of such services. In a wide number of occasion though, private information are captured by online services without the direct user consent or even knowledge. We believe that the privacy and sensitiveness of the information becoming accessible to third parties can be easily overlooked. 

Personal computers and more generally communication devices that are carried around by people are capable of being located, identified and tracked across different locations, networks and services~\cite{michael2013location}. All these devices can therefore be used for a variety of surveillance activities, which are in itself detrimental to the user's interests. Until recently in fact, the cost of surveillance and tracking of people and activities was proportional to the cost of directly reaching, asking or following a single person or a group of people. Technology therefore enhances the surveillance capabilities by introducing tools that allow the collection of information arising from a person's activities. This information can furthermore be combined and inferred, therefore offering a more complete picture of that person. 

For example, to personalise their services or offer tailored advertising, web applications could use tracking services that identify a user through different networks~\cite{veeningen2014line}~\cite{getoor2012entity}. These tracking services usually combine information from different profiles that users create, for example their Gmail address or their Facebook or LinkedIn accounts. In addition specific characteristics of the user's device can be used to identify them through different sessions and websites, as described by the Panopticlick project~\cite{eckersley2011panopticlick}.

Browser fingerprinting is a technique implemented by analytics services and tracking technologies to identify uniquely a user while they browser different websites. Different features of a specific browser setup can be used to identify uniquely a user. Supported languages, browser extensions or installed fonts~\cite{boda2012user} can be used to identify a browser setup among others. More advanced techniques distinguish between browsers' JavaScript execution characteristics~\cite{mowery2011fingerprinting}. These features are particularly interesting since they are more difficult to simulate or mitigate in practice. Targeting JavaScript execution characteristics actually means looking at the innate performance signature of each browser's JavaScript engine, allowing the detection of browser version, operating system and microarchitecture. These attacks can also work in situations where traditional forms of system identification (such as the user-agent header) are modified or hidden. Other techniques exploit the whitelist mechanism of the popular NoScript Firefox extension.This mechanism allow the user to selectively enabling web pages' scripting privileges to increase privacy by allowing a site to determine if particular domains exist in a user's NoScript whitelist.

It is important to note that while tracking creates serious privacy concerns for internet users, the customisation of results is also beneficial to the end user~\cite{castelluccia2012behavioural}. In fact, while tailored services offer to the user only information relevant to their interests, it also allows some companies and institutions to concentrate an enormous amount of information about internet users in general. ~\cite{rao2015they} investigate user profiling and access mechanisms offered by online data aggregator to users' collected data. Both the collected data and its accuracy was analysed together with the user's concerns. In their findings about 70\% of the participants to the study expressed some concerns about the collection of sensitive data, its level of detail and how it might be used by third parties, especially for credit and health information.

It has been shown how most successful tracking networks exhibit a consistent structure across markets, with a dominant connected component that, on average, includes 92.8\% of network vertices and 99.8\% of the connecting edges~\cite{gomer2013network}. ~\cite{gomer2013network} have measured the chance that a user will become tracked by all top 10 trackers in approximately 30 clicks on search results to be of 99.5\%. More interesting, ~\cite{gomer2013network} have shown how tracking networks present properties of the small world networks. Therefore implying a high-level global and local efficiency in spreading the user information and delivering targeted ads.

An interesting property of networks to understand their architecture is the behaviour of the average degree of nearest neighbours~\cite{barrat2004architecture}~\cite{pastor2001dynamical}. The average degree of the nearest neighbours of a node $k_{nn}(k)$ is a quantity related to the correlations between the degree of connected vertices ~\cite{maslov2002specificity}, since it can be expressed as the conditional probability that a given vertex with degree $k$ is connected to a vertex of degree $k'$. This property defines if the network in consideration is assortative, if $k_{nn}$ is an increasing function of k or dissortative ~\cite{newman2002assortative} if it is not. The property of assortativity has been used  in the field of epidemiology, to help understand how a disease or cure spreads across a network. It is particularly interesting to note that assortativity can give a measurement if the removal of a set of network's vertices may correspond in curing, vaccinating or quarantining individuals cells in the network.

Protection techniques against tracking networks are implemented through software agents able to identify if third-party requests are accessing private data. These agents include Privacy Badger ~\cite{privacy-badger}, Mozilla Lightbeam ~\cite{lightbeam}, Ghostery ~\cite{ghostery}, AdBlock ~\cite{adblock}, and so on. Some of these agents implement a Tracking Protection Lists (TPL). A TLP can be seen as a blacklist of identified tracking domains that user might want to block. 

Another interesting aspect of advertising services is how they are designed to work on feedback loops \cite{degeling2016your}. An advertising service can in fact be seen as a blackbox providing the tracker trying to identify or profile the user, and the returned advertising content. The tracker is used to send information back to the advertising service, which in response will return a certain content tailored to the user preferences. Within this feedback loop different aspect of the user behaviour are taken in consideration. These include certainly the users browsing history and their click through rate, i.e. a measurement of the amount of time users in a population are more likely to interact with an ad. In more sophisticated advertising solution also user social connections are taken in consideration.

Advertising therefore services raise the problem of confidentiality of the user reading activity ~\cite{ard2013confidentiality}. Up to know an eloquent example of this problem was provided by the way public library in the US operates. Reading activities were considered historically private and were protected through a set of rules that restricted  libraries ability to exploit reading records.This regime is clearly bypassed when libraries decide to provide digital services to their users. Digital services providers and third parties can in fact access users reading activities without agreeing to the library confidentiality regime.

\section{Modelling the user's footprint}

We model the user's activity as series of events belonging to a certain identity. Each event is a document containing different information. We can formally defined this as a hypermedia document i.e. an object possibly containing graphics, audio, video, plain text and hyperlinks. We call the hyperlinks selectors and we use these to build the connections between the user's different identities or events. Each identity can be a profile that the user has created onto a service or platform, or just a collection of events, revealing something about the user. With account we mean an application account or a social network account, such as their LinkedIn or Facebook unique IDs. An event is an action performed by the user, like visiting a website.

\begin{figure}[!ht]
\centering
\includegraphics[scale=0.45]{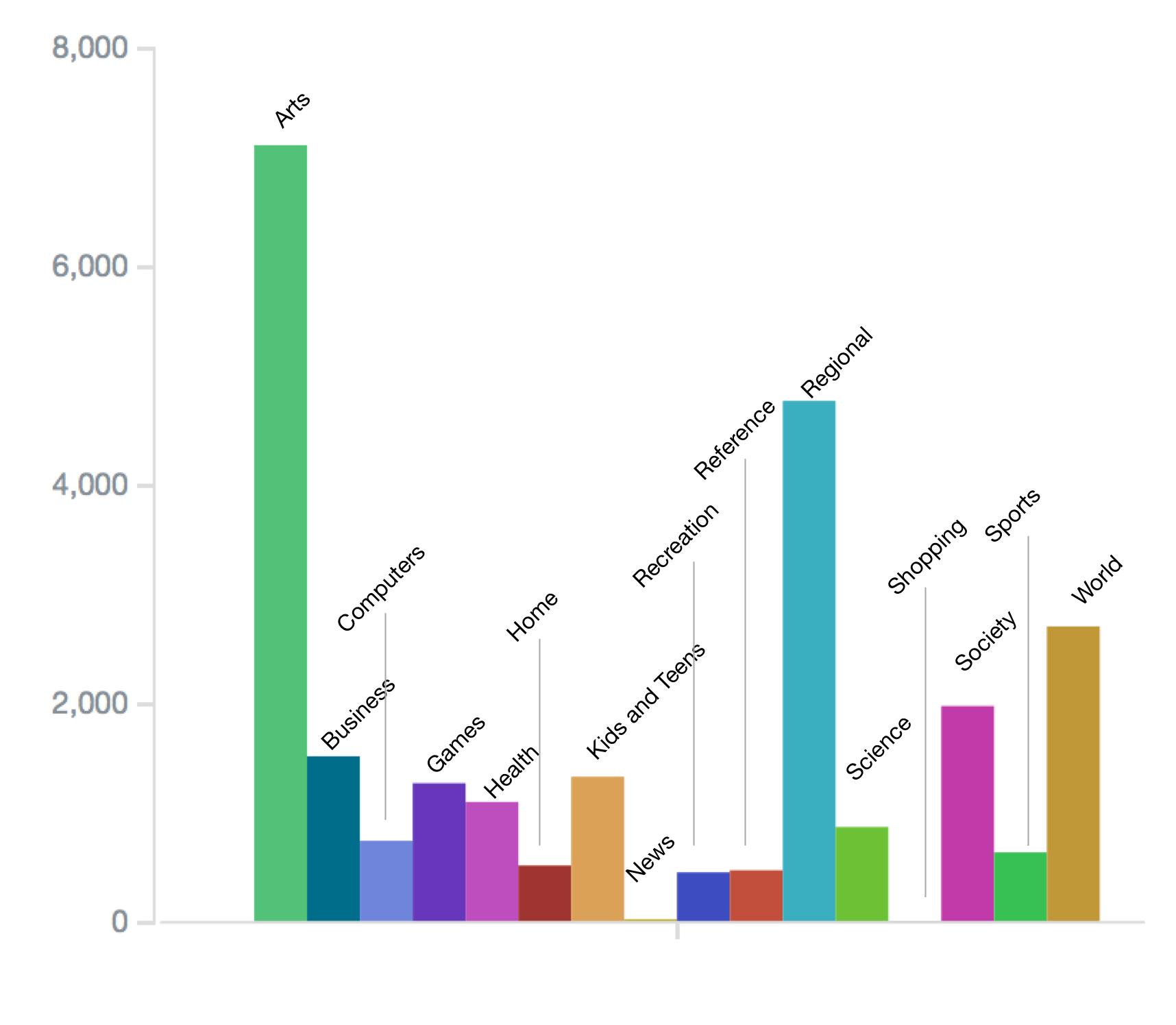}
\caption{Here we show an example of user profile expressed in absolute terms by counting the number of keywords in each category for a browsing session. We model user and advertising profiles as histograms of tags keywords a set of predefined categories of interest.
\label{abs-profile}}
\end{figure}

We aggregate keywords each time the user creates a new event by visiting a different url. These keywords constitute the user profile of interests (Figure ~\ref{abs-profile}). A tractable model of the user profile as a probability mass function (PMF) is proposed in~\cite{Parra12DKE,Parra12TKDE} to express how each keyword contributes to expose how many times the user has indirectly expressed a preference toward a specific category. We consider that the user expresses a preference when they visit a webpage categorised with a certain keywords. This model follows the intuitive assumption that a particular category is weighted according to the number of times this has been counted in the user  profile.

We define the profile of a user $u_m$ as the PMF $p_m = (p_{m,1},\ldots, p_{m,L})$, conceptually a histogram of relative frequencies of tags across the set of tag categories $\mathcal{T}$.

Similarly, we define the profile of an ads, or third-party http request, $i_n$ as the PMF $q_n =(q_{n,1},\ldots, q_{n,L})$, where $q_{n,l}$ is the percentage of tags belonging to the category $l$ which have been assigned to this specific advertising item. 

Both user and ads profiles can then be seen as normalised histograms of tags across categories of interest. Our profile model is in this extent equivalent to the tag clouds that numerous collaborative tagging services use to visualise which tags are being posted, collaboratively or individually by each user. A tag cloud, similarly to a histogram, is a visual representation in which tags are weighted according to their relevance.

In view of the assumptions described in the previous section, our privacy attacker boils down to an entity that aims to profile users by representing their interests in the form of normalised histograms, on the basis of a given categorisation.

\subsection{Third party requests on web pages}

\begin{figure}[!ht]
\centering
\includegraphics[scale=0.45]{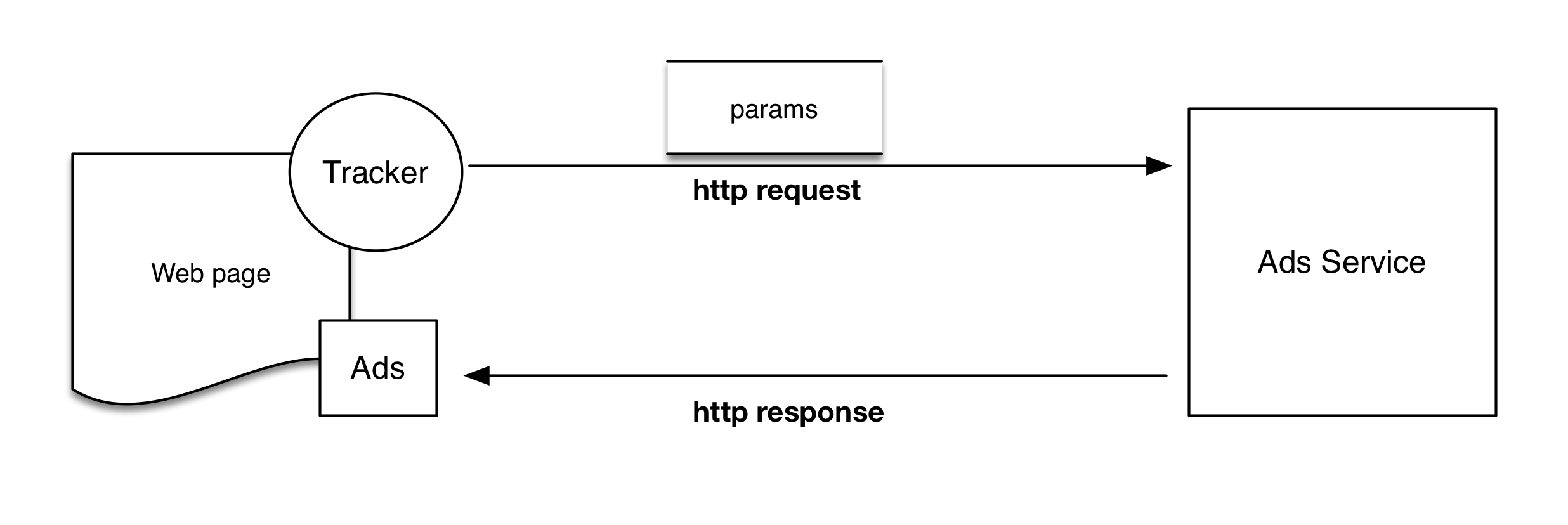}
\caption{Trackers on web pages make third-party http requests to advertising services. These return ads content tailored to the user web history or expressed preferences.
\label{third-party}}
\end{figure}

When a user visits a web page, the browser sends an http request to the server to request a representation of the resource described through the url. The server provides the resource representation in the form of a html document and the browser parses it. The html document contains a number of links to other resources, such as JavaScript code, videos, audios or images (Figure ~\ref{third-party}). Some of these can be stored on the same domain as the requested page, some may be requested to a third party services. Such is the case of services like Google Analytics, share buttons from different social networks, or advertising banners. Together with the http request, a number of parameters are included. These contains keywords, users preferences, information regarding the user device and session, in page information sent to the third party service from the website or application.

\subsection{A metric of similarity}

We consider the third party advertising network to operate like a recommendation system, that suggest products or services that might be of interest for the user, based on their preferences. A recommendation system can be described as an information filtering system that seeks to predict if the user is interested or not in a particular resource. We assume that the ad server suggest advertising based on a measure of \emph{similarity}. 

We measure the user profile, as previously described, as an histogram of their recorded preferences, and the advertising profile as an histogram of the ads that the user has received. We have considered a set of metric to measure how the advertising network is tracking the user profile. We use the \emph{1-norm} as a first measure of \emph{distance} between the advertising profile and the user profile:

$$ \| p_{m} - q_{n} \|_1 = \sum_l{ | {p_m}_l - {q_n}_l | } $$

The \emph{1-norm} is related to the total variation distance commonly used in statistics when considered over a finite alphabet. In this case, given two probability distributions, P and Q, over a finite alphabet, we can relate the total variation distance to the \emph{1-norm} as follows:

$$ \delta(P, Q) = \frac {1} {2}  \| P - Q \|_1 =  \frac {1} {2} \sum_l{ | P(l) - Q(l) | } $$

We also use the \emph{2-norm} as measurements of the distance between the advertising network and the user profile:

$$ \| p_{m} - q_{n} \|_2 = (\sum_l{ {| {p_m}_l - {q_n}_l |}^2 })^{1/2} $$

The \emph{2-norm} represents the Euclidean distance between the two distributions. When considering the \emph{2-norm} it is possible to highlight larger discrepancies on the set of categories analysed.

\begin{figure}[!ht]
\centering
\includegraphics[scale=0.40]{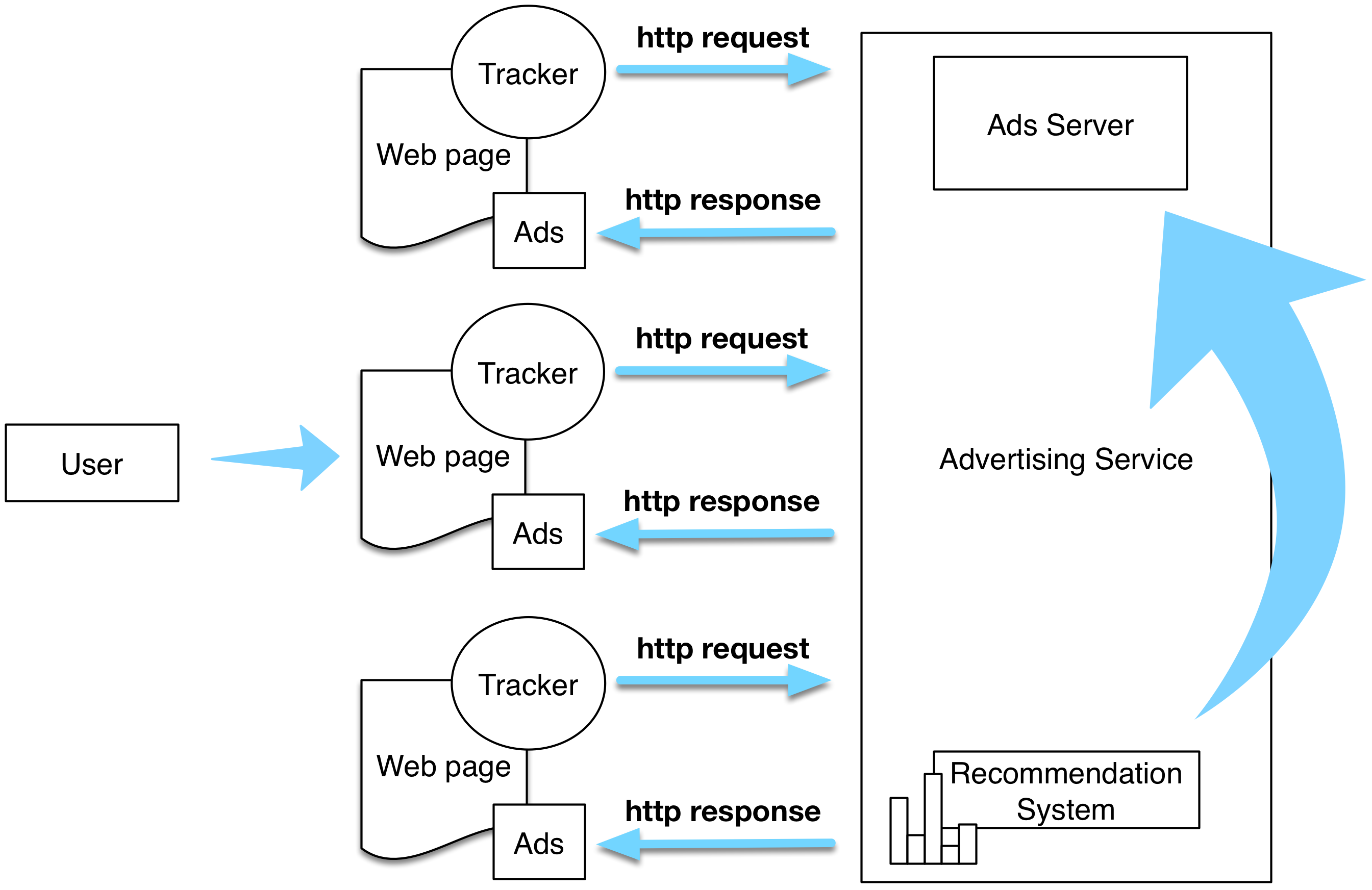}
\caption{Advertising services work in a feedback loop. The image illustrate how while a user surf a number of web pages, the service record their profile and adapts the returned advertising.
\label{advertising-loop}}
\end{figure}

Advertising services are complex recommendation systems working in a feedback loop (Figure ~\ref{advertising-loop}). When a user surfs the web each tracker on the visited pages communicates with the advertising service, sending a number of parameters through http requests. These contain the user preferences and browsing history which will be taken into consideration when ads are returned to display on the page. 

\subsection{A metric of connectivity}

We said that advertising networks or privacy attackers need to be able to \emph{follow} the user across as many website as possible in order to profile their interests. We have therefore considered that the average degree of the neighbourhood of each node is a good indication of how many pages are connected to a certain advertising service or tracking domain.

The average degree of the neighbourhood of a node $i$ is calculated as:

$$ k_{nn,i} = \frac{1}{| N(i) |} \sum_{j \in N(i) } {k_j} $$

Where $N(i)$ are the neighbours of node $i$ and $k_j$ is the degree of node $j$ which belongs to $N(i)$.

If a certain tracker domain is connected to the majority of the page visited by a certain user, this means that they have been able to collect the user's preferences and reading activities across a number of websites. The more a tracker domain is connected, the more the user might consider this a \emph{risk} for their privacy.

We have therefore used the average degree of the neighbourhood of a tracker to rank tracker domains. 

\section{Experimental methodology and results}

We analysed the browsing habits of 86 users of Twitter, by observing the set of 10 websites links shared for each of the top categories from the Open Directory Project (DMOZ)~\cite{a22}. We assumed that the articles shared on twitter are a subset of the website that each users visit every day. More importantly if they are active Twitter users, these websites will express their interest bias towards certain categories. To validate our strategy we observed that Twitter itself offer website owners the possibility to track conversions on their pages coming from tweets and twitter ads. Please note that the list of links was only considered as a list of website visited, no interaction between Twitter user was further taken into consideration. 

These sites are therefore surfed in our simulation environment. This consist of a virtual box were a browser instance visits a url and record both in page categories and third-party requests. In this scenario we pretend that a user is going through their reading list of sites and by looking at third-party http requests we measure how the advertising changes in the page and adapts to their profile. The user is simulated by a software agent opening the urls and scrolling through the page for a certain arbitrary amount of time.

It is important to note that in our simulated environment the users are not logged a third-party account, like Google, Facebook or Twitter. When the website is accessed a text version of the page is recorded and analysed by our software agent. In page keywords and meta information are extracted and evaluated. We extracted keywords from the actual text of the page by using the Rapid Automatic Keyword Extraction (RAKE)~\cite{RAKE} algorithm. Each keyword was then evaluated against Open Directory Project (DMOZ)~\cite{a22} for classification within top levels categories.

Once the user profile was calculated the advertising profile is evaluated. The advertising profile is extracted from url parameters contained in third party requests. We have collected information regarding each third party requests made from each page visits. These parameters are again evaluated against DMOZ for classification within top levels categories. Please note that we have excluded request made to JavaScript libraries, images and Cascade Style Sheet (CSS) files. We have also excluded same domain requests, since we were only interested in third-parties http calls.

By profiling users' browsing events using a hypermedia document structure we were able to show how each event contains a set of features regarding the user identity and the page that was visited. We have therefore categorised each event by using the keywords contained in the meta information present in the page and the page text itself (Figure ~\ref{abs-profile}). At each event we ere able to calculate an event profile, by measuring the set of keywords introduced by each action performed by the user (i.e. visited a page).

For each user we considered a series of 15 pages visited and we measured how the norm (both \emph{1-norm} and \emph{2-norm}) between the measured advertising profile and the user profile changed at each visit. We considered this strategy to follow the intuition that advertising is probably tailored on historical data up to the current page visit. Therefore new page visits need a certain amount of time to be \emph{counted}.

\begin{figure}[!ht]
\centering
\includegraphics[scale=1.00]{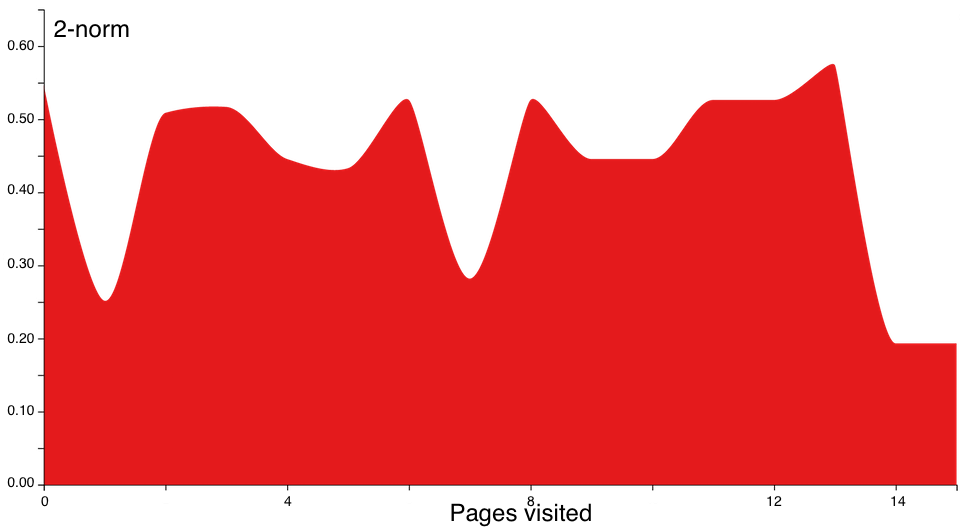}
\caption{The image show the \emph{2-norm} for a single user visiting a series of 15 pages.
\label{user-2-norm}}
\end{figure}

In this scenario, we expected the norm to fluctuate and quickly adapt to the user calculated profile. For this reasons when evaluating the distance between the advertising profile and the user profile we always considered the measured user profile up to the current page visited (Figure ~\ref{user-2-norm}).

\begin{figure}[!ht]
\centering
\includegraphics[scale=0.20]{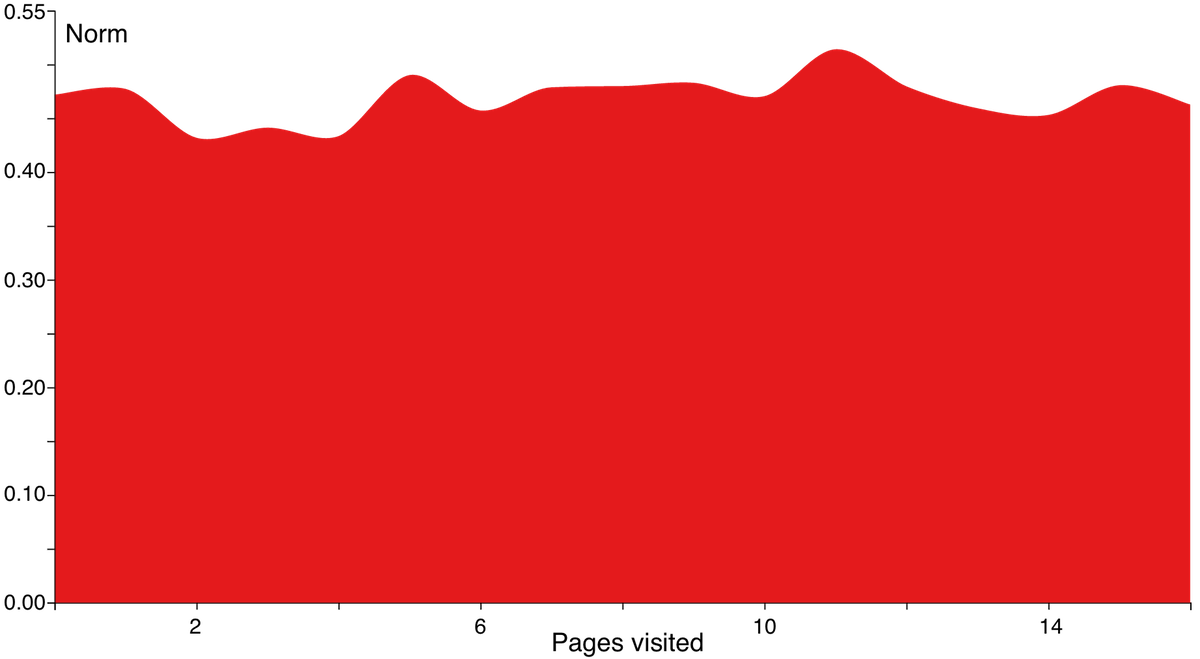}
\caption{The image show the average \emph{1-norm} for all the users in the experiment visiting a series of 15 pages.
\label{pop-1-norm}}
\end{figure}

\begin{figure}[!ht]
\centering
\includegraphics[scale=0.80]{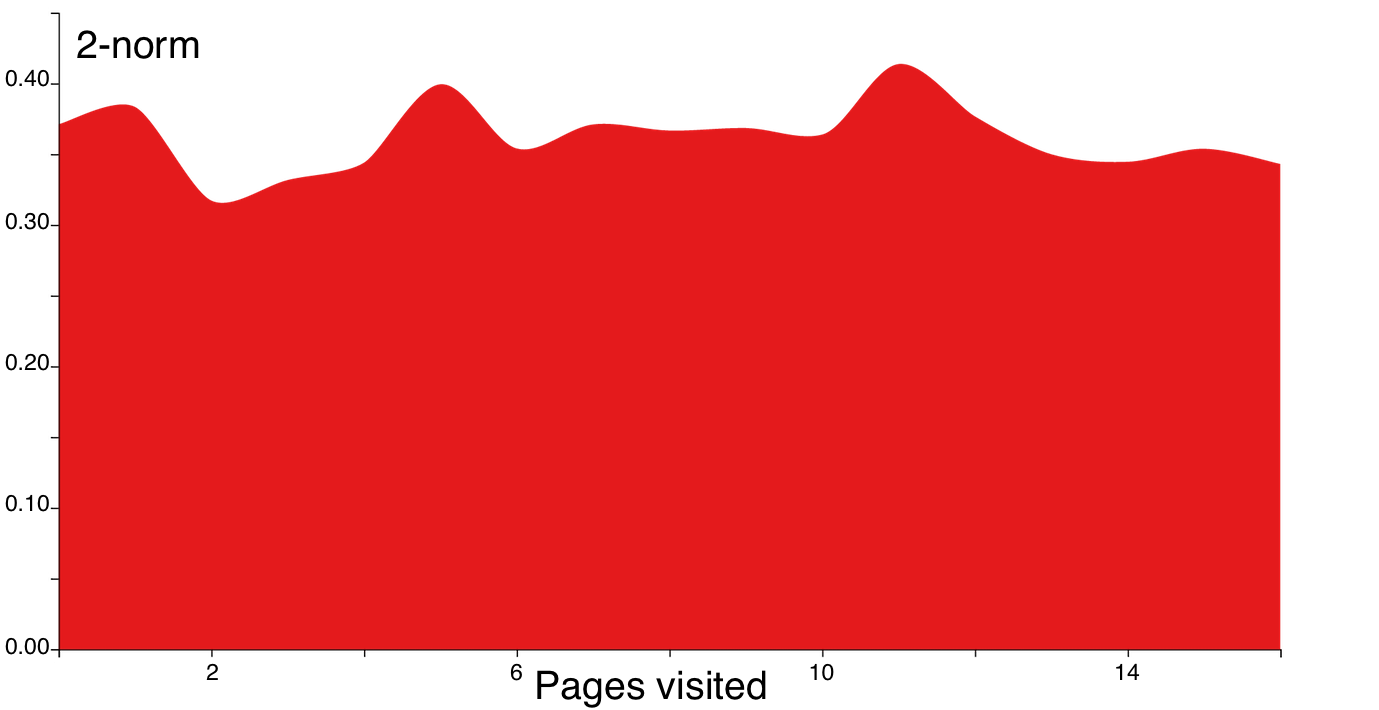}
\caption{The image show the average \emph{2-norm} for all the users in the experiment visiting a series of 15 pages.
\label{pop-2-norm}}
\end{figure}

We also considered a average measurement of how the norm changes for the whole population of users evaluated, for a series of 15 pages each (Figure ~\ref{pop-1-norm} and ~\ref{pop-2-norm}).

\begin{figure}[!ht]
\centering
\includegraphics[scale=0.45]{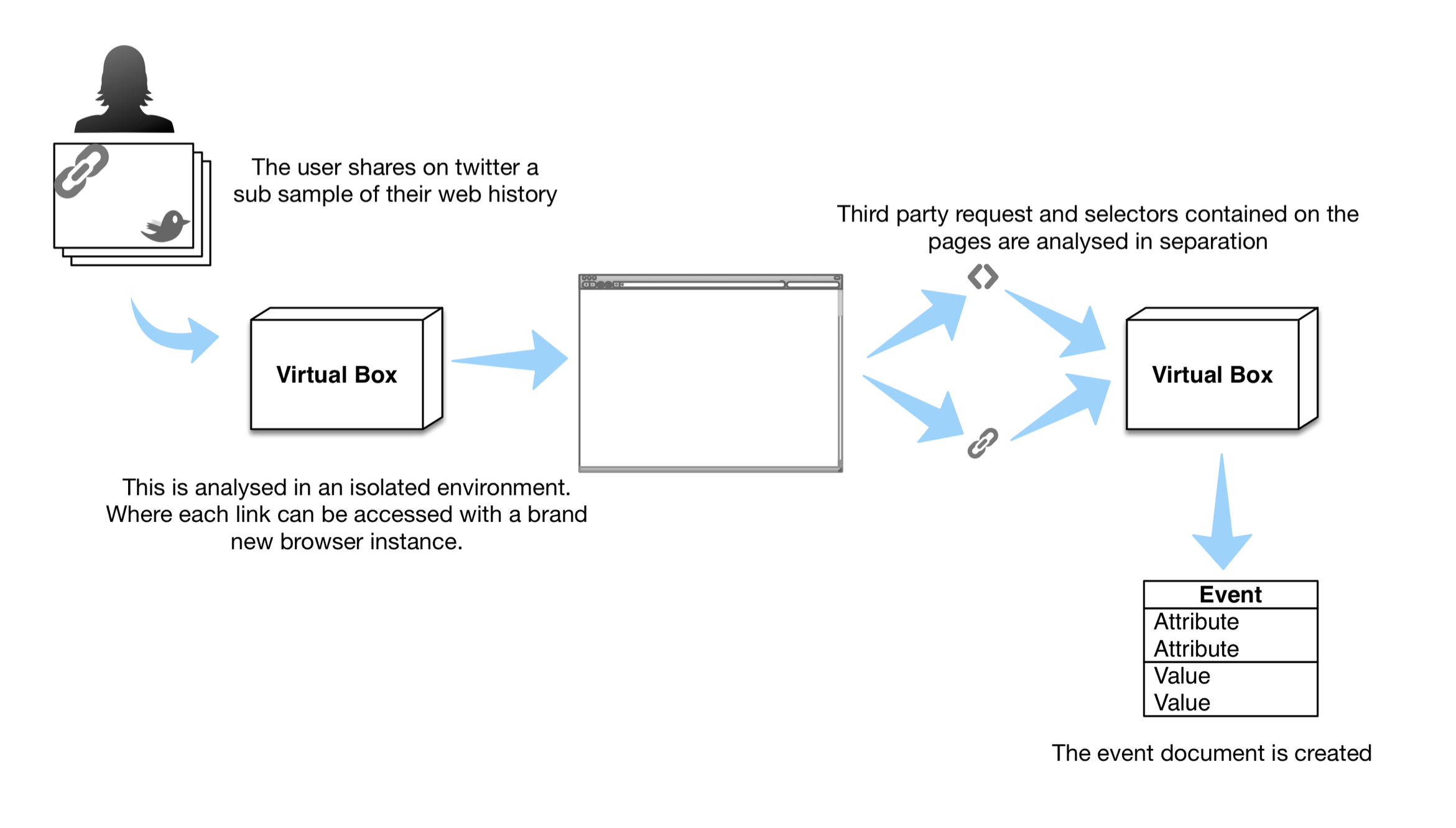}
\caption{The image show how Twitter feeds were used to construct users' browsing histories to use in our experiment.
\label{meth}}
\end{figure}

It is interesting to note that in a short number of visits the norm measurements for the entire population of users decrease of approximately $5\%$ for the \emph{1-norm} and $3\%$ for the  \emph{2-norm}. For an individual user we see instead how the decrement in the \emph{2-norm} is more evident ~\ref{user-2-norm} with a decrement of approximately $35\%$ in 15 pages visited.

By using our hypermedia model we also considered how tracker domains are linked to pages. In this case we calculated the average degree of the neighbourhood of each node, for nodes corresponding to advertising services. Our results shows how we were able to identify known tracker domains.

\begin{table}[h!]				
\centering																												
\begin{tabular}{ || c | c || } 
\hline
Tracker domain & avg $k_{nn,i}$ \\
\hline
\hline
tacoda.at.atwola.com & 180.0 \\
bcp.crwdcntrl.net & 180.0 \\
match.prod.bidr.io & 180.0 \\
glitter.services.disqus.com & 180.0 \\
ad.afy11.net & 180.0 \\
idsync.rlcdn.com &180.0 \\
mpp.vindicosuite.com & 180.0 \\
aka-cdn-ns.adtechus.com & 180.0 \\
clients6.google.com & 180.0 \\
i.simpli.fi & 180.0 \\
ads.p161.net' & 180.0 \\
dis.criteo.com & 180.0 \\
ads.stickyadstv.com & 180.0 \\
cms.quantserve.com & 180.0 \\
ads.yahoo.com & 129.0 \\
graph.facebook.com & 118.0 \\
ib.adnxs.com & 110.0 \\
rs.gwallet.com & 108.0 \\
bid.g.doubleclick.net & 98.333\\
googleads4.g.doubleclick.net & 98.333\\
\hline
\end{tabular}
\label{table-trackers}
\\[2.5pt]
\caption{The table shows the top 20 identified tracker domains based on the average degree of the neighbourhood.}
\end{table}

\section{Conclusions and future work}

We measured how the norm adapt quickly to the user profile in a short amount of pages visited to expose how advertising networks are able to profile users even when they are not logged into an identity account. This means that the networks possess a large amount of information regarding users and population of users to be able to predict fairly quickly user's preferences. This aspect is particularly interesting for the development of Privacy Enhancing Technologies for the web. Up to now, anti-tracking technologies have been built to simply stop third party requests, alternative strategies might instead consider to send bogus information to certain over-connected tracker domains to masquerade the user real profile. At the same time a measurement of the average degree of the neighbourhood of a certain third-party domain can be used to evaluate how \emph{dangerous} this can be considered for the user's privacy.
In future research we would like to further explore the hypermedia model introduced, while continuing to understand how quickly web advertising is able to match the served ads with the actual user profile. This would allow us to understand if different profiles for the same users can be somehow linked together within similar advertising networks. We are also particularly interested in measuring how social networks sharing buttons and/or commenting services, included on websites, are able to track users even when these have not signed in  with  their  account.  We  reserve  the  study  of  their  capabilities  to  future  investigations. More over we want to enlarge the set of users analysed by testing on logs from a real world small computer network, while also introducing new metrics to our study. In particular we are already planning to consider: the KL-Divergence  between  the  advertising  profile  and  the  observed  user  profile. We also believe in the importance to provide users with simple visualisation tools able to show the user their online footprint and allowing them to take action to masquerade their interests profile or simply block certain networks.

\section*{Acknowledgment}

This work was supported by the Spanish Ministry of Economy and Competitiveness (MINECO) through the "Anonymized Demographic Surveys (ADS)" project, ref. TIN2014-58259-JIN, under the funding program "Proyectos de I+D+i para Jóvenes Investigadores", and through the project "INRISCO", ref. TEC2014-54335-C4-1-R, as well as by the Government of Catalonia, under grant 2014 SGR 1504. 



%\bibliographystyle{IEEEtran}

\bibliographystyle{IEEEtran}

\bibliography{Bibliography/StringAbbreviated,Bibliography/Security,Bibliography/InfoTheory,Bibliography/LosslessCoding,Bibliography/LossyCoding,Bibliography/MathStatSigPro,Bibliography/Classification,Bibliography/Applications,Bibliography/ReferencesTRIPP,Bibliography/SemanticWeb,Bibliography/InsubriaReferences,Bibliography/rfc,Bibliography/Silvia_bibliography,Bibliography/thesis_proposal}

\end{document}